\documentclass
[twocolumn,prb,showpacs,amsmath,amstex,amssymb,mathfonts,superscriptaddress]{revtex4}
\usepackage{amsthm,amsfonts,graphicx,verbatim}
\usepackage{amsmath}
\usepackage{amssymb}
\usepackage{amsthm}
\usepackage{amsfonts}
\usepackage{listings}
\usepackage{enumerate}
\usepackage{latexsym}
\usepackage{psfrag}
\usepackage{bm}
\usepackage{graphicx}
\usepackage{subfigure}
\newcommand{\be}{\begin{equation}}
\newcommand{\ee}{\end{equation}}
\newcommand{\bea}{\begin{eqnarray}}
\newcommand{\eea}{\end{eqnarray}}

\newcommand{\la}{\langle}
\newcommand{\ra}{\rangle}

\renewcommand{\phi}{\varphi}
\renewcommand{\epsilon}{\varepsilon}
\renewcommand{\vec}[1]{{\bf #1}}

\newcommand{\sket}[1]{|#1\rangle}

\newcommand{\ex}[1]{\left \langle #1 \right \rangle}

\begin{document}
\title{Fractional and integer  quantum Hall effects in the zeroth Landau level in  graphene}
\author{Dmitry A. Abanin} 
\affiliation{Department of Physics, Harvard University, Cambridge,
Massachusetts 02138, USA}
\affiliation{Perimeter Institute for Theoretical Physics, Waterloo, Ontario N2L 6B9, Canada}
\affiliation{Institute for Quantum Computing, Waterloo, Ontario N2L 3G1, Canada}
\author{Benjamin E. Feldman}
\affiliation{Department of Physics, Harvard University, Cambridge,
Massachusetts 02138, USA}
\author{Amir Yacoby}
\affiliation{Department of Physics, Harvard University, Cambridge,
Massachusetts 02138, USA}
\author{Bertrand I. Halperin}
\affiliation{Department of Physics, Harvard University, Cambridge,
Massachusetts 02138, USA}

\date{\today}
\begin{abstract}
Experiments on the fractional quantized Hall effect in the zeroth Landau level of graphene have revealed some striking differences between filling factors in the ranges $0<|\nu|<1$ and $1<|\nu|<2$.   We argue that these differences can be largely understood as a consequence of the effects of terms in the Hamiltonian which break  $SU(2)$ valley symmetry, which we find to be important for $|\nu|<1$ but negligible for $|\nu| >1$.  The effective absence of valley anisotropy for $|\nu|>1$ means that states with odd numerator, such as $|\nu|=5/3$  or 7/5 can accommodate charged excitations in the form of large radius valley skyrmions, which should have a low energy cost, and may be easily induced by coupling to impurities. The absence of observed quantum Hall states at these fractions is likely due to the effects of valley skyrmions.  For  $|\nu|<1$, the anisotropy terms favor phases in which electrons occupy states with opposite spins, similar to the antiferromagnetic state previously hypothesized to be the ground state at $\nu=0$. The anisotropy and Zeeman energies suppress large-area skyrmions, so that quantized Hall states can be observable at a set of fractions similar to those in GaAs two-dimensional electron systems. In a perpendicular magnetic field $B$, the competition between the Coulomb energy, which varies as $B^{1/2}$, and the Zeeman energy, which varies as $B$, can explain the observation of apparent phase transitions as a function of $B$ for fixed $\nu$, as transitions between states with different degrees of spin polarization. In addition to an analysis of various fractional states from this point of view, and an examination of the effects of disorder on valley skyrmions, we present new experimental data for the energy gaps at integer fillings $\nu=0$ and $\nu= -1$, as a function of magnetic field, and we examine the possibility that valley-skyrmions may account for the  smaller energy gaps observed at $\nu= -1$.
\end{abstract}
\maketitle

%
%
%
%
%
%
%

\section{Introduction}

The Dirac spectrum and high four-fold valley and spin degeneracy of Landau levels (LLs) in graphene lead to interesting phenomena in the quantum Hall effect (QHE) regime~\cite{CastroNeto09}. In particular, Coulomb interactions lift LL degeneracies, giving rise to new integer quantum Hall states with broken spin and/or valley symmetry~\cite{Zhang06,Checkelsky07,Young12,Nomura06,Abanin06,Alicea06,Yang06,Goerbig06,FertigBrey06,Abanin07,Gusynin08,Jung09,DungHai,Chamon,Kharitonov11}. Some of these states were conjectured to have counter-propagating spin-filtered edge states~\cite{Abanin06,FertigBrey06,Gusynin08}, as well as unusual spectra of collective excitations~\cite{Yang06}. 

Recently, the progress in sample fabrication led to the observation of fractional quantum Hall (FQH) states in high-mobility graphene samples (suspended and graphene on BN)~\cite{Du09,Bolotin09,Dean11,Feldman12,Smet12,Feldman13}, in both transport measurements, and in measurements of the local incompressibility using a scanning single-electron transistor (SET) technique.  Recent experimental study~\cite{Feldman12} revealed an unexpected sequence of FQH states in the zeroth LL. Certain FQH states in the interval $-2<\nu <-1$ of filling factors were missing or very weak, including states at filling factors $\nu=-\frac{5}{3}, -\frac{7}{5}, -\frac{11}{7}$. This is quite surprising, because fractional states with denominator $3$ are generally expected to be strong. In the interval of filling factors $-1<\nu < 0$, the FQH sequence was essentially identical to that in found in GaAs-based 2D electron system~\cite{prange}. More recently, incompressibility measurements  in the range $-1 <\nu < 0$ have revealed transitions, as a function of the applied perpendicular magnetic field,  at a fixed filling factor, which have been interpreted as transitions between  states with different spin polarizations\cite{Feldman13}. It is worth noting that the absence of a $\nu=-5/3$ plateau was first observed in experiment~\cite{Dean11}. Also, new multi-component FQH states arising due to valley 
and spin degeneracy of LLs in graphene were considered theoretically~\cite{multi-component}. 

Here we propose some possible explanations for the puzzling observations made in Refs.~[\onlinecite{Dean11,Feldman12,Feldman13}].
We argue that the key differences between the first partially filled Landau sub-level ($-2<\nu<-1$) and the second one ($-1<\nu<0$) arise from the greater importance  in the second regime of the valley anisotropy terms in the Hamiltonian, which break the SU(2) symmetry with respect to the valley index.  

In what follows, we will use both the carrier filling factor $\nu \equiv n \Phi_0 /B_{\perp}$,  where $n$ is the net carrier density, $B_{\perp}$ is the magnetic field perpendicular to the sheet and $\Phi_0$ is the flux quantum, and the ``Landau level filling factor" $f\equiv \nu+2$, to  describe filling of the zeroth LL. (When there are no carriers present, $\nu=0$, the LL is half full, and  $f=2$).  We shall concentrate on filling factors $-2<\nu<0$, as  the filling factors $0<\nu<2$ are related to these by particle-hole symmetry. 

The rest of the paper is organized as follows. In Section~\ref{hamiltonian} we introduce the Hamiltonian and discuss how particle-hole symmetry can be used to relate the energies of different states. In Section~\ref{integer_states} we review the known results about integer states at filling factors $\nu=-1$ and 0, and argue that the excitations of these states are different, which makes the energy gap at $\nu=-1$ smaller than the gap at $\nu=0$. In Section~\ref{fqhe1} we study various FQH states in the filling factor interval $-2\leq\nu\leq-1$. We argue that the states at odd-numerator fractions generally have skyrmion excitations, while the excitations of the even-numerator states are quasi-particles and quasi-holes. In Section~\ref{fqhe2} we turn to FQH states at $-1<\nu<0$. We discuss the valley and spin structure of the states, finding that at all filling factors in this interval except for $\nu=-1/3$ there are two or more competing states with different degree of valley and spin imbalance. The competition between Coulomb energy and Zeeman energy results in multiple phase transitions that occur as a function of magnetic field. In Section~\ref{disorder} we consider the effects of disorder on various fractional and integer states, and argue that disorder can weaken and even suppress the odd-numerator states at filling factors $-2<\nu<-1$. Section~\ref{skyrmions} contains new experimental data on the energy gaps of the $\nu=-1$ and $\nu=0$ states, as well as a discussion about the nature of the excitations of these states. We summarize our findings in Section~\ref{summary} and compare them to recent experiments~\cite{Dean11,Feldman12,Feldman13}. Finally, in Section~\ref{outlook} we mention possible future directions for studies of fractional quantum Hall effect in graphene and related materials. 
 
\section{Projected Hamiltonian}
\label{hamiltonian}

To understand the underlying symmetries, let us start with the microscopic Hamiltonian of the zeroth LL. We will neglect LL mixing, and consider projections of all operators onto the zeroth LL. 
For the analysis of spin and valley effects, it is convenient to separate the Hamiltonian into the leading $SU(4)$-symmetric part $\hat H_{\rm sym}$ and terms that break spin and valley symmetry.  The symmetric part may be written as
\be\label{eq:H_symm}
\hat{H}_{\rm sym}=\frac{1}{2}\int d{\bf r} d{\bf r'}: \hat\psi^\dagger({\bf r}) \hat\psi({\bf r}) V({\bf r-r'})
\hat\psi^\dagger({\bf r'}) \hat\psi({\bf r'}): ,  
\ee
where the  symmetric interaction is given, at least approximately, by a Coulomb potential,  $V({\bf r-r'})=\frac{e^2}{\epsilon |{\bf r-r'}|}$, where $\epsilon$ is an effective dielectric constant. Also, $\hat\psi^\dagger ({\bf r})$ here is an operator that creates an electron in the zeroth LL. Since there are four fermion species, we view it as a four-component field, 
\be\label{eq:psi}
\hat\psi^\dagger ({\bf r})=\left(\hat\psi^\dagger_{K\uparrow} ({\bf r}), \hat\psi^\dagger_{K\downarrow} ({\bf r}), \hat\psi^\dagger_{K'\uparrow} ({\bf r}), \hat\psi^\dagger_{K'\downarrow} ({\bf r})\right)
\ee
The four components  may be regarded as forming a spinor in a four-dimensional space, which we shall refer to as ``hyperspin space," and we shall refer to an arbitrary  vector in this space as a ``hyperspin" state. 

There are three main symmetry-breaking terms. First, the Zeeman interaction breaks spin rotation symmetry: 
\be\label{eq:Zeeman}
\hat H_{Z}=E_Z \int d{\bf r} \hat\psi^\dagger ({\bf r}) \hat\sigma_z \hat\psi({\bf r}), \,\, E_Z=\frac{g}{2}\mu_B B
\ee
The most important valley-anisotropic terms arise due to lattice-scale Coulomb interactions~\cite{Alicea06,Jung09,Kharitonov11}, as well as due to electron-phonon interactions~\cite{DungHai,Chamon}. There are two types of valley-anisotropic terms, which can be written as follows, 
\be\label{eq:H_anis_z}
\hat H_{\rm an}^z=\frac{g_z}{2}\int d{\bf r} : \left[  \hat\psi^\dagger({\bf r}) \hat \tau_z \hat\psi({\bf r}) \right]^2 :
\ee
\be\label{eq:H_anis_p}
\hat H_{\rm an}^{\perp}=\frac{g_{\perp}}{2}\int d{\bf r} \sum_{i=x,y} : \left[  \hat\psi^\dagger({\bf r}) \hat \tau_i \hat\psi({\bf r}) \right]^2 :  
\ee
The coupling constant $g_z$ originates from the electron-electron interactions, while $g_{\perp}$ includes contributions from both electron-electron and electron-phonon interactions~\cite{Alicea06,DungHai,Chamon,Kharitonov11}. The valley-anisotropic terms can get renormalized and enhanced by LL mixing~\cite{Jung09,Kharitonov11}, but here we will treat them as phenomenological parameters.  We also assume that the interactions can be treated as instantaneous in
time. We note that LL mixing will also lead to some renormalization of the  SU(4) invariant  Coulomb interaction, on the scale of the magnetic length, but we ignore such effects here.

The valley anisotropy terms in our model, (\ref {eq:H_anis_z}) and   (\ref {eq:H_anis_p}), have effect only when two electrons coincide in space.  This reflects the fact that on the microscopic scale, sensivity to the electron valley or sublattice occurs only when two electron are separated by a distance of the order of the graphene lattice constant, which is much smaller than the magnetic length.  We also remark that in the zeroth Landau level, the electrons in opposite valleys are confined to opposite sublattices. 

With these assumptions  the contributions of $g_z$ and  $g_{\perp}$ to the energy per flux quantum,  for a fixed value of $\nu$,  will scale as $g_z / (2 \pi l_B^2) $ and  $g_{\perp}    / (2 \pi l_B^2) $, respectively.  Here $l_B=\sqrt{\hbar c/eB_\perp}$ is the magnetic length. (We will mostly consider the case when magnetic field is perpendicular to the sample, therefore, in the rest of the paper we identify $B$ with $B_\perp$). Consequently, this energy will scale the same way as the Zeeman energy, proportional to $B$, as previously remarked.  (We note that the relative importance of these terms can be altered, however, by application of a parallel field, which will increase the Zeeman energy but not affect the valley anisotropy energies.)   Because the energies due to Coulomb interactions scale as $B^{1/2}$, we see that  the ratio between these and the  valley anisotropy and  Zeeman terms will change when the magnetic field is varied at fixed filling factor, and transitions between different ground state configurations could occur as a result. 

\subsection{Implications of electron-hole symmetry} 

As was mentioned in the Introduction, the Hamiltonian described above, when projected onto a single Landau level,  has an exact particle hole symmetry.  Specifically, it can be shown that given any many-body state $\sket {\Phi}$ with LL filling factor $f$ and energy $E_\Phi$, there exists a particle-hole conjugate state $\sket {\Psi}$, with LL filling factor $4-f$, whose energy is given by  
\be
E_\Psi = E_\Phi + \frac {f-2}{2} E_4 \, ,
\ee
where $E_4$ is the energy of the completely full Landau level, i.e., the state with $f=4$. The above expression can be obtained by straightforward generalization of the analysis for the case of two Landau sub-levels performed in Ref.~[\onlinecite{Fertig94}]. 

For the subsequent analysis, it is also helpful to note that there are additional electron-hole symmetries if we restrict the hyperspin states occupied by the electrons.  For example, if we only consider many body states with electrons belonging to a single, fixed, hyperspin state, there is a particle-hole symmetry relating a state  $\sket {\Phi}$ with LL filling factor $f$ and a state $\sket {\Psi}$ with  LL filling factor $1-f$, such that  $E_\Psi = E_\Phi +( 2f-1) E_1$.  
If we consider many-body states with electrons belonging to two specified orthogonal hyperspin states, then there is a particle-hole symmetry relating a state  $\sket {\Phi}$ with LL filling factor $f$ and a state $\sket {\Psi}$,  with LL filling factor $2-f$, such that  
\be
\label{ph2}
E_\Psi = E_\Phi +( f-1) E_2 .
\ee  
We will use these relations below when discussing the energies of various fractional states.

\section{IQH States}\label{integer_states}

\subsection{Nature of the ground state.}

We first consider the ground states at $f=1$ and $f=2$, which is to say at $\nu = -1$ and $\nu = 0$.  If the terms in the Hamiltonian which break the SU(4) symmetry are set equal to zero, the ground-state at $f=1$ is predicted to be simple Slater determinant state, in which we have selected an arbitrary state $\sket{\chi}$ in the four-dimensional hyperspin vector space, and filled all orbital states in the Landau level with the given hyperspin.  Similarly, at $f=2$, we expect the ground state to consist of two filled Landau levels with arbitrary orthogonal hyperspin states $\sket{\chi_1}$ and $ \sket{\chi_2 }$.  (More precisely, it can be shown that these Slater determinants are exact ground states of the SU(4)-symmetric Hamiltonian for a wide range of interaction potentials, which almost certainly includes the standard Coulomb interaction. We note, however, that there could be lower energy states for more peculiar interactions.)
The energies of these states are  given precisely by the Hartree-Fock approximation.

As long as  the Hamiltonian is chosen to be SU(4) invariant, the ground state energies of the Slater determinant states will be independent of the choice of the occupied hyperspin states. The valley anisotropy  terms  and the Zeeman energy lift the ground state degeneracy, at least partially.  To find the true ground state or ground states, one should calculate the expectation value of the anisotropy terms for  arbitrary choice of the  occupied hyperspin states, and choose a state which minimizes the total energy.  

For the case of $f=1$, the energy  minimization is quite simple. When only a single hyperspin state is occupied, there will be zero probability density for two electrons to sit at the same point in space because of the Pauli exclusion principle.  Since the valley anisotropy terms in our Hamiltonian act non-trivially only if two electrons occupy the same point in space, their expectation value will be zero in any wave function that contains  only a single occupied hyperspin state .  Therefore, splitting of the SU(4) degeneracy can only arise from the Zeeman term. The ground state will then have electron spin aligned with the applied magnetic field, but the valley pseudospin state $\sket{v}$ may point anywhere on the Bloch sphere. 

For $f=2$, the valley anisotropy terms come into play, because the Pauli principle does not prevent electrons in different hyperspin states from  coinciding in space.  In this case ground states with several different symmetries are possible, depending on the signs and relative sizes of  $g_z$, $g_{\perp}$, and the Zeeman energy. These states have been classified and their mean-field phase diagram is summarized in Ref.~[\onlinecite{Kharitonov11}].  If the Zeeman term were large compared to the valley anisotropy terms, electrons in the lowest LL at $\nu=0$ would have their spins completely aligned with the magnetic field, and the ground state would thus be a valley singlet.   However, if this were the case, the state at $\nu =0$ should have a finite electrical conductance, due to the contribution of zero-energy energy edge states~\cite{Abanin06,FertigBrey06}, which is contrary to experimental  results~\cite{Checkelsky07,Du09,nu01}.  One possible state, which would have an energy gap at the edge as well as in the bulk is the  Kekule state~\cite{DungHai,Chamon}, which is a spin-singlet state in which all electrons occupy  a single linear combination of the two  valley states, with 50 \%  probability to be in either valley. Another possibility, which, however, seems unlikely in the presence of strong short-range repulsion, would be  a ``charge-denstiy-wave'' state where electrons occupy both spin states in a single valley. This state was discussed in Refs.~[\onlinecite{Alicea06,Jung09}].   

Another possibility, explored by Jung and MacDonald~\cite{Jung09}, as well as by Kharitonov~\cite{Kharitonov11}, is an antiferromagnetic (AF)  state in which electrons have spin in a direction  $\hat{s_1}$ on one sublattice (say sublattice $A$) and have a different orientation $\hat{s_2}$ on the second sublattice. (Recall that for electrons in the lowest LL, electrons in  valleys $K$ and $K'$  will reside on different graphene sublattices.) In the absence of a Zeeman field, the orientations $\hat{s_1}$ and $\hat{s_2}$ will point opposite to each other, but can lie along an arbitrary axis in space. In the presence of a weak Zeeman field, the spins will cant slightly and line up so that their sum $\hat{s}_1+\hat{s}_2$ points in the direction of the magnetic field $\vec{B}$, while
$\hat{s}_1-\hat{s}_2$ lies in the plane perpendicular to $\vec{B}$. As noted by Jung and Macdonald~\cite{Jung09} and subsequently by Kharitonov~\cite{Kharitonov11}, this canted antiferromagnetic  (CAF) state could transform continuously to a spin-aligned state in the presence of a sufficiently large parallel magnetic field. The AF state will be the ground state, in the absence of a Zeeman term, if   $g_z > 0,\,  g_{\perp}<0$, and  $g_z > |g_{\perp}|$.  The expectation value of the anisotropy energy in the AF state,  per flux quantum,  is equal to $E_2^{\rm{an}} = - g_z / (2 \pi l_B^2)$.  For small Zeeman coupling, the canting angle will be small, and the energy gain due to canting will be equal to $\pi l_B^2 E_Z^2 / |g_{\perp}|$. 

Current experimental evidence points to the CAF state being the ground state at $\nu=0$ in the absence of parallel field component~\cite{Young_private}. At certain large parallel field, the insulating CAF state gives way to the metallic FM state~\cite{Young_private}, similar to the case of bilayer graphene~\cite{Maher12}. To characterize the relative strength of the valley anisotropy and Zeeman interactions, we introduce a quantity
\be\label{eq:gamma}
\gamma \equiv \frac{2\pi l_B^2 E_Z} {  |g_{\perp}|}. 
\ee
Based on experimental evidence, including the behavior of transport coefficients at $\nu=0$  in a  tilted magnetic field, it appears that $\gamma$ is of order 0.1~\cite{Young_private}. Therefore, the Zeeman energy is small compared to the valley anisotropy energy at $f=2$, and the canting angle is small. 

In our analysis of integer and fractional states below, we will consider parameter values that favor CAF state ($g_z>0, g_\perp<0$, and $g_z>|g_\perp|$), and also mostly assume that the Zeeman interaction is much smaller than the valley anisotropy.

 \subsection{Charged excitations at $f=1$.} 

Given our characterizations of the ground states at $f=1$ and $f=2$, it is natural to ask about the elementary charged excitations in the two cases.  As we shall see below, the nature of excitations is quite different in these two cases: at $f=1$, the excitations are skyrmions, while  at $f=2$ the excitations are electrons and holes. 
 
We first consider the case of $f=1$. The simplest model for a hole excitation is the state in which one simply removes one electron from the filled level with the chosen hyperspin state $\sket{\chi}$. The simplest model for an electron excitation would be to add a single electron in one of the hyperspin states orthogonal to $\sket{\chi}.$ The energy of a hole excitation is given by 
$$
E_h=\frac{\Delta_0}{2}+E_Z, \,\, \Delta_0=\sqrt{\frac\pi 2}\frac{e^2}{\epsilon l_B}, 
$$
irrespective of its hyperspin. Quantity $\Delta_0$ is the energy gap for creating an electron-hole pair in the absence of symmetry-breaking terms. 
In the case of graphene, the energy of an electron excitation, however, contains a contribution from valley-anisotropic terms in the Hamiltonian, and therefore is sensitive to the hyperspin of the extra electron. Let us assume that the ground state at $f=1$ is polarized along $K\uparrow$ direction in the hyperspin space. Then we can consider extra electrons with hyperspin $K'\downarrow$ or $K'\uparrow$. Their respective energies are given by
$$
E_{e,AF}=\frac{\Delta_0}{2}-\frac{g_z}{2\pi l_B^2}+E_Z, \,\, E_{e,FM}=\frac{\Delta_0}{2}-\frac{2g_\perp+g_z}{2\pi l_B^2}-E_Z. 
$$
We also note that adding extra electron with $K\downarrow$ will cost more energy. Given that $\frac{|g_\perp|}{2\pi l_B^2}\gg E_Z$, the electron excitation with hyperspin $K'\downarrow$ ("AF" one) has the lowest energy. 

For Coulomb interactions, however,we expect that  the lowest energy charged excitations at $f=1$ will  be skyrmions, in which the hyperspin is misaligned relative to $\sket {\chi}$ over a large area~\cite{Sondhi93,Fertig94} (for a discussion of skyrmions in SU(N)-symmetric quantum Hall ferromagnet, see Ref.~[\onlinecite{Arovas99}]). In the absence of symmetry breaking terms, Coulomb interactions will favor skyrmions of infinite radius, where the skyrmion energy is found to be one-half of the energy of a single particle electron or hole excitation~\cite{Sondhi93,Fertig94}. In GaAs, the radius of the area of the skyrmion, and the number of overturned spins, is limited by the Zeeman energy, which gives an energy cost proportional to the skyrmion area~\cite{Sondhi93}.

In graphene, hyperspin can take 4 different values, so several kinds of skyrmions are possible. One possibility is to consider valley skyrmions -- textures in which valley polarization is smoothy varied but spin is aligned with the magnetic field.  There is no Zeeman cost for such a valley skyrmion, and we would expect that the size of an isolated skyrmion in an ideal sample would be arbitrarily large, to minimize the Coulomb interactions.  (In practice, the size of a skyrmion might  be limited by disorder or by the distance between skyrmions).  In any case, the energy of a valley skyrmion should be close to 1/2 of that of a single electron or hole excitation: 
\be
\label{g1}
E_{sk}=\frac{\Delta_0}{4}\pm E_Z+E_v ,
\ee
where $E_v$ denotes the contribution of valley-anisotropic terms, and $\pm$ refers to hole(electron) skyrmions.


To examine the effect of valley anisotropy terms, first note that a large area skymion may be characterized by a space dependent orientation  in hyperspin space for the occupied electron state.  At the center of the skyrmion, the hyperspin should point in a direction determined by a spinor $\sket{\chi_2}$  that is orthogonal to the state $\sket{\chi}$ which characterizes the background far from the skyrmion.  
We define a normalized  probability  density  $P$ for two electrons to coincide at a point $\vec{r}$,  by 
\be
\label{Pr}
P(\vec{r}) = (2 \pi l_B^2)^2  \ex { : \hat{\psi}^\dagger(\vec{r})    \hat{\psi}(\vec{r}) \hat{\psi}^\dagger(\vec{r}) \hat{\psi}(\vec{r})  :   }  
\ee 
Because the valley anisotropy terms are relevant only when two electrons coincide in space,  any contribution of these terms to the energy density of the skyrmion must be proportional to $P(\vec{r})$. (We note that two electrons with valley indices $K,K'$ in the zeroth LL can never reside on the same lattice site, because the zeroth LL wave functions in the $K(K')$ valleys reside solely on the $A(B)$ sublattice. So, by "coincide in space" we mean that two electrons are much closer than the magnetic length $l_B$.) A priori, we would expect that $P$ will be of order $l_B^2 /R^2$ inside the skyrmion of radius $R$ (one extra electron spread out over area $~R^2$) and that it will fall to zero outside.  

A deeper analysis of the skyrmion structure at $f=1$ leads to the conclusion that $P(\vec{r}) =0$ in the case of the hole skyrmion~\cite{MacDonald96}, at least in the limit of large skyrmion radius.  Therefore there is no anisotropy energy for the hole skyrmion, and 
$$
E_{sk,h}=\frac{\Delta_0}4+E_Z. 
$$
For the electron skyrmion, $P \neq 0$, since we have added an extra electron.  Furthermore, it follows from particle-hole symmetry that   $\int P(\vec{r}) \, d^2r=1$ for the electron skyrmion. Therefore, there will be a non-zero anisotropy energy in this case, equal to $E_v=-\frac{2g_\perp+g_z}{2\pi l_B^2}$. This gives the energy of the spin-aligned electron valley skyrmion: 
\be\label{eq:skyrmion_FM}
E_{sk,e}^{FM}=\frac{\Delta_0}4-E_Z-\frac{2g_\perp+g_z}{2\pi l_B^2}.  
\ee
This is the  energy in the limit of infinite skyrmion size. Smaller skyrmions  will be more costly due to the Coulomb interactions. 

As an alternative to the valley skyrmion, one may consider a spin-involved skyrmion, where $\chi_2$ has a spin orientation opposite to the magnetic field. In this case,  $\sket{\chi_2}$ may have a valley pseudospin  that is the same or is different from that of  $\sket{\chi}$.  Taking into account the implications of Eq.(\ref{ph2}), we find that if the lowest energy state at $f=2$ is the AF state, the electron skyrmion at $f=1$ with lowest anisotropy energy will be one where the hyperspin state far from the skyrmion has electrons in one valley and spin aligned with the magnetic field, while at the center of the skyrmion the electrons  sit in the opposite valley and have reversed spin.  The anisotropy energy in this case would be  $ E^{\rm{an}}_ {\rm{skyrm},2} = -  (2 \pi l_B^2)^{-1}  (g_z ) $, just like for the single electron-like excitation.

The total energy of the ``AF" skyrmion is given by:
\be\label{eq:AFskyrmion}
E_{sk,e}^{AF}=\frac{\Delta_0}{4}-\frac{g_z}{2\pi l_B^2}+(2 n_s-1)E_Z  +D\frac{e^2}{\epsilon R}, 
\ee
 where $R$ is the nominal  skyrmion radius, the term $De^2/\epsilon R$ arises from the long-range Coulomb interaction energy ($D$ is a numerical constant of order unity), and  $n_s(R)$, the number of electrons with reversed spins,  is proportional to $(R/ l_B)^2$, multiplied by a constant that depends logarithmically on $R$ and $E_Z$.  The Zeeman interaction favors small skyrmions, while the Coulomb interaction favors infinitely large skyrmions. The competition of these two terms leads to an optimal intermediate-size skyrmion 
 with energy~\cite{Sondhi93,Bychkov}
 \be\label{eq:optimal_sk}
 E_{sk,e}^{AF}=\frac{\Delta_0}{4}-\frac{g_z}{2\pi l_B^2} + A \Delta_0   \tilde g_Z^{1/3} |\log  (10 \, \tilde g_Z)|^{1/3} \,  .
 \ee
 Here
$$
 A=\frac{3^{4/3} \pi^{5/6}}{2^{25/6}}\approx 0.62, 
$$ 
and
 $$
 \tilde g_Z=\frac{E_Z}{e^2/\epsilon l_B}
 $$
 is the Zeeman interaction in units of Coulomb interaction energy (notice that the dimensionless Zeeman interaction differs from the definition adopted in Refs.[\onlinecite{Sondhi93,Bychkov}] by a factor of 4). The expression (\ref{eq:optimal_sk}) is valid in the limit $\tilde g_Z\ll 1$.   The number of electrons with reversed spins may be found from
 $$
(2 n_s - 1)  =      \frac{d E^{AF}_{sk,e}}{ dE_Z} \approx  \frac{A}{3}  \sqrt{\frac\pi 2} \,  \tilde g_Z^{-2/3} |\log  (10 \, \tilde g_Z)|^{1/3}
$$
The difference of the energies of two kinds of skyrmions is given by:
 \be\label{eq:en_diff}
 \Delta E_{sk,e}=E_{sk,e}^{AF}-E_{sk,e}^{FM}=A\Delta_0  \tilde g_Z^{1/3} |\log (10 \tilde g_Z)|^{1/3}+\frac{2|g_\perp|}{2\pi l_B^2}. 
 \ee
 The first term scales as $B^{2/3}$ with magnetic field, while the second term scales as $B$. Therefore, at low fields we expect the large valley skyrmions to be the lowest-energy excitations, while at high fields, spin-reversed skyrmions will have lower energy. The critical field at which the nature of the skyrmions changes is found from the condition $\Delta E_{sk,e}=0$, which can be rewritten as follows, 
 $$
 A\sqrt{\frac\pi 2} |\tilde g_Z|^{-2/3}|\log  \tilde g_Z|^{1/3}=\frac{2}{\gamma}. 
 $$
 This gives the critical dimensionless $g$-factor (with logarithmic precision):
 $$
 \tilde g_Z\approx \left(\frac{A\gamma}{2}\sqrt{\frac{\pi}2} \right)^{3/2} \log^{1/2}\left[10 \left(\frac{A\gamma}{2}\sqrt{\frac{\pi}2} \right)^{3/2} \right]
 $$
 For $\gamma=0.1$, and $\epsilon = 5.24$ (the RPA result for intrinsic screening function of graphene~\cite{screening_function}), this yields 
 $$  \tilde g_Z\approx 0.0075, \,\,
 B_*\approx 1.8\, {\rm T}. $$ 
 Thus, at $B>B_*$ we expect the lowest-energy excitations to be spin-reversed skyrmions. 
 
 The number of reversed spins for AF skyrmions can be accurately determined using numerical results of Ref.[\onlinecite{Fertig97,Cooper97}] (the effective sigma-model result is valid only in the limit when parameter $\tilde g_Z$ is very small, which is not necessarily the case for experimentally relevant fields $B\sim 10\, {\rm T}$). Within a classical (i.e., Hartree-Fock) description,  the mean number of electrons with reversed spins $n_s$ in a skyrmion will  vary continuously with the radius of the skyrmion. In a proper quantum description, however,
the skyrmion must be an eigenstate of the spin angular momentum parallel to the  applied magnetic field, and hence the value of $n_s$ will be
quantized and  have only  integer values.  Similarly, the number of electrons $n_v$ with the reversed valley index $KÕ$ must also be an
integer. Different types of skyrmions are thus properly characterized by specifying the two integers $n_s$ and $n_v$.  For the AF skyrmion
described above, we have $n_v=n_s$, while for the pure valley skyrmion, we have $n_s=0, n_v = \infty$.  We may also consider skyrmions in the
quantum equivalent of a canted AF state, which would have  $0 < n_s < n_v < \infty$.  However, it is not clear that such skyrmions would be favored in the parameter range of interest. 
 
 For magnetic fields $B=2.5, 5, 10\, {\rm T}$ we find, using results of Ref.[\onlinecite{Cooper97}], $n_s\approx 5, 4, 3$ for AF skyrmions. The number of reversed spins can be measured in parallel field experiments. Recent transport experiments on samples on BN substrate found that the activation gaps at $f=1$ were sensitive to the parallel field component, indicating that they involve spin flips~\cite{Young12}. The effective $g$-factor was found to be somewhat enhanced, which is consistent with small spin skyrmion excitations. It should be noted, however, that the spin-reversed and valley skyrmions might have different sensitivity to disorder, and their transport properties might be different as well. Therefore it is possible that large valley skyrmions are present and contribute to compressibility, but are more easily pinned by disorder and therefore their contribution to transport is suppressed. Future compressibility measurements in tilted field might shed light on the nature of excitations at $f=1$.

\subsection{Charged excitations at $f=2$.}

In an SU(4)-symmetric model, skyrmions are possible at filling fraction $f=2$, and have the same energy as at $f=1$. In the case of graphene, however, the short-range valley-anisotropic terms presumably favor an AF state, and, in the presence of Zeeman energy, the symmetry is broken down to $Z(2)\times U(1)$ (where $Z(2)$ is the valley symmetry and $U(1)$ is the symmetry related to the rotation of the spin orientation in the $xy$-plane). Because of the low symmetry, skyrmions cost a large anisotropy energy proportional to their area, in contrast to the case of $f=1$. As a result, skyrmions are not favorable and the lowest excitations are single-particle-like. We expect that this would make the energy gaps at $f=2$ significantly larger than the gaps at $f=1$. 

Let us suppose that the ground state at $\nu=0$ is the AF state, with occupied states $K\uparrow$ and $K'\downarrow$. 
A single hole excitation, in which we remove an electron with spin that is anti-aligned with magnetic field (hyperspin $K'\downarrow$), has an energy 
$$
E_h=\Delta_0-{E_Z}+\frac{g_z}{2\pi l_B^2},  
$$
while an electron excitation with hyperspin $K'\uparrow$ (to minimize Zeeman interaction) has an energy
$$
E_e=\Delta_0-{E_Z}-\frac{2g_\perp}{2\pi l_B^2} 
$$
The corresponding activation gap is then given by
\be\label{eq:gap_2}
\Delta_{\nu=0}=E_h+E_e=2\Delta_0-2 E_Z+\frac{g_z-2g_\perp}{2\pi l_B^2}. 
\ee

Turning to the possibility of having skyrmions at $\nu=0$, we note that several types of charged skyrmions are possible. 
A charged skyrmion might have a filled LL for $K \uparrow$ and a valley-pseudospin texture which mixes $K' \downarrow$ and $K\downarrow$. At the center of the skyrmion the system would be locally in a CDW state, with both spin states in valley $A$, which would cost an energy proportional to $g_z$ and to the area of the skyrmion.  Similarly, we could consider a texture that mixes $K'\downarrow$ and $K'\uparrow$. Such a configuration would cost an energy proportional to  $|g_{\perp}|$, multiplied by the skyrmion area. Therefore, the energy of the charged skyrmion involves a valley anisotropy cost proportional to the skyrmion area. 

Given that the valley anisotropy is large, we therefore expect the skyrmions at $f=2$ to be unfavorable. The relevant parameter that determines the size of the skyrmions and their energy is $\tilde{g}=\frac{(|g_\perp|/2\pi l_B^2)}{e^2/\epsilon l_B}$. Taking $\gamma=0.1$, we obtain $\tilde{g}\approx 0.35$, which is well above the critical values $\tilde g\approx 0.06$ at which skyrmions become completely suppressed and lowest energy excitations are electrons/holes~\cite{Fertig97,Cooper97}. Thus we conclude that the lowest-energy excitations at $f=0$ are single holes and electrons, with the corresponding gap given by Eq.(\ref{eq:gap_2}).

\section{FQH States for $ 0 < f <1$}\label{fqhe1}

We start our analysis of the FQH effect in graphene by considering various states in the interval of filling fractions $0<f<1$. 

\subsection{ $f=1/3\,\,\,\,(\nu = -5/3)$}

The ground state at  $f=1/3$ is presumably well described by the Laughlin trial wave function, with the  electrons in a single hyperspin state, with spin aligned by the magnetic field and arbitrary orientation of the valley pseudospin.   The Laughlin state is the exact ground state in the limit of a zero-range interaction potential, and is known to be very accurate for a Coulomb potential.  Moreover, the assumption that the ground state contains only one hyperspin state is consistent, as the valley-anisotropy terms have no effect on such a state due to the Pauli exclusion principle, and the state has been already  chosen to minimize the Zeeman energy.  

The lowest energy charged excitations at $f=1/3$ are probably valley skyrmions. The size of such a skyrmion can be very large, as it does not cost any Zeeman energy, in contrast with the case of a conventional semiconductor, where the skyrmion involves spin orientations that are not aligned with the magnetic field.    As far as we are aware, there are currently no reliable calculations of the spin-stiffness constant for an $f=1/3$ state with Coulomb interactions, so it is difficult to estimate the size of the energy gap that would be produced.  However, based on numerical calculations for small systems~\cite{Wojs02,Toke06} it is expected that the skyrmion energy will be significantly smaller than the energy of a Laughlin quasiparticle or quasihole.  

Unlike the skyrmions at $f=1$, the excitations at  $f=1/3$ should have only a very small probability of two electrons being at the same position in space, for quasi-electron-like as well as a quasi-hole-like skyrmions. Indeed, it is possible to construct a trial wave function for a pseudospin-reversed electron, or for a skyrmion,  which costs energy due to the $V_2$ pseudopotential, but has no contribution from $V_0$, and it is probable that even for Coulomb interactions, the skyrmion wave function will tend to strongly  avoid such a contribution~\cite{MacDonald98,Jain_skyrmion}. Consequently, the valley-anisotropic interactions should  have very little effect on the energy of the skyrmions.

\subsection{$ f=2/5\,\,\,\,(\nu = -8/5)$}

We expect that the ground state at $f=2/5$ should be a pseudospin singlet state, with two valley states equally occupied and with electron spins aligned with the magnetic field.  The elementary trial wave function for this state is the Halperin 332  wave function~\cite{Halperin83}, in which the amplitude vanishes as the cube of the separation when two electrons in the same valley state approach each other, and vanishes quadratically for electrons in opposite valleys.  This is an exact ground state for a system of  $N \geq 2$  species of particles with an  SU(N)-invariant Hamiltonian in the limit of a zero-range repulsive interaction, since it is the only wave function at filling 2/5 in the lowest LL which as no contributions from the $V_0$ and $V_1$ pseudopotentials.  It is also found to be an excellent approximation in the case of a Coulomb potential. 

 In GaAs, the spin-singlet state is in competition with a spin-aligned 2/5 state that has a higher Coulomb energy but is favored by the Zeeman term, so there exists a transition between the two  states  as one varies the magnetic field at fixed filling factor, or if one increases the Zeeman energy by application of a parallel magnetic field.   In graphene,  however, the spins are already fully aligned with the magnetic field in the valley-singlet state, so there is no Zeeman  penalty  for forming the state.  Therefore, we predict that a state where all electrons have the same hyperspin configuration should never be the ground state in graphene at $f=2/5$, and there should be no transition as one varies the magnetic field.

\subsection{$ f=2/3 \,\,\,\,(\nu = -4/3)$}

At first glance, there are two plausible candidates for the state at $f=2/3$. One may consider a state in which all electrons have the same hyperspin configuration, or one may consider a state which is a valley-singlet but spins aligned by the magnetic field~\cite{Apalkov06}, as for the $f=2/5$ state discussed above.  In GaAs, at $f=2/3$, it is found that  the spin singlet state has a lower Coulomb energy than the spin-aligned state, but as at $f=2/5$, the spin-aligned state is favored by the Zeeman energy,  and there is a transition between the two states depending on the ratio of the Zeeman energy to the Coulomb energy energy scale.  In graphene, by contrast, the valley singlet and the hyperspin-aligned states would have the same Zeeman energy, so we would expect the valley singlet to be the ground state at all magnetic fields.  However,   we should also consider the possible contribution of the anisotropy terms before concluding that this is the case. 

As noted previously,  a state with all electrons in the same hyperspin state will have no energy contribution arising from the valley anisotropy terms, due to the Pauli exclusion principle.  On the other hand, the valley singlet state at $f=2/3$ could have a non-zero probability density $P$ for two electrons in opposite valleys to coincide at the same point in space.   [We continue to use  the normalization of $P$
defined by (\ref{Pr}), which is such that $P=1$ in the AF state at $f=2$.] 
 Then if $P_{2/3}$ is the value of $P$ in the $f=2/3$ valley-singlet state,  the anisotropy energy, per flux quantum  will be  
$-(g_z+ 2 g_{\perp}) P_{2/3} / (2\pi l_B^2) $, which will be positive if 
 $( g_z + 2 g_{\perp}) < 0 $.  This inequality is consistent with the condition $ \ g_z > - g_{\perp} >0 $  for stability of the AF state at $\nu = 0$, but is not required by it.  
As previously noted, the anisotropy energy per electron at fixed $f$ will scale linearly with the applied magnetic field, just as for the Zeeman energy in GaAs, so if  $( g_z + 2 g_{\perp}) >0 $, it is conceivable, in principle, that there could be a transition between the two types of ground states in graphene at $f=2/3$.  However, it appears that the value of $P_{2/3}$ is actually extremely small, so that the valley singlet is always the ground state.  

Unlike the case of the 332 wave function at $f=2/5$, there is no simple trial wave function  for the valley-singlet state at $f=2/3$ where one can evaluate the value of $P$. Nevertheless, numerical calculations on a finite system suggest that  $P_{2/3}$ has a value less than $10^{-4}$ for pure Coulomb interactions~\cite{morf13}.  This is, in fact,  very  much smaller than the 
 value $P=1/9$  that one would obtain if  
 one filled one third of the available states in each valley with spin parallel to the magnetic field, and assumed that there were no spatial correlations between electrons in opposite valleys.
The small value of $P_{2/3}$ also guarantees that any advantage in anisotropy energy could never outweigh the Zeeman penalty for a state in which spins are not all aligned with the magnetic field.
If we add a  skyrmion to the valley singlet state,  it will necessarily cost a substantial Zeeman energy, so we do not expect there to be low-energy charged excitations in the form of large-radius skyrmions.

We note that if the ground state at $f=2/3$ were actually  the state with electrons confined to a single hyperspin state, it would be a particle hole conjugate of the state at $f=1/3$.   Just as for the $f=1/3$ state, there would be low-energy skyrmion excitations, which mix in the second valley state but do not  cost any Zeeman energy.  

Within the language of composite fermions,  the valley-singlet state at $f=2/3$ would be described as a state where one has doubly filled the lowest composite fermion LL, in an effective magnetic field that is opposite in sign to the applied field,  with particles in two orthogonal hyperspin states.  Alternatively, it is possible  to describe the valley-singlet  2/3 state as one that can be obtained by adding a 1/3 density of hole-like charged excitations to the fully polarized state at $f=1$. In this case, the charged excitations would be minimally-sized hole-like skyrmions, which may be described as  a bound state of two missing electrons in the occupied hyperspin state and one added electron in the reversed valley state. As the difference in occupation numbers of the two valley states is decreased by three for each added hole, the difference will be reduced to zero at $f=2/3$. 

\subsection {$ f=4/5 \,\,{\rm {and}} \,\, 6/7 \,\,\, (\nu = - 6/5, \, \, -8/7)$}

We may consder several possible candidates  for a state at $f=4/5$.  One possibility would be a completely hyperspin polarized state, in which all electrons occupied the same valley state and had spin aligned with the magnetic field.  Such a state would be a particle-hole conjugate of the state at $f=1/5$, and would be described as a Laughlin 1/5 state of holes in the full hyperspin-polarized Landau level. This state would presumably be the correct ground state in a model of short-range interactions, with a dominant $V_0$ repulsive  pseudopotential, but it is unlikely to be the true ground state in the case of Coulomb interactions.  

As was previously noted, at $f=1$, the lowest energy hole excitation,  in the case of Coulomb interactions, should be a large-area skyrmion with many reversed valley spins, rather than a single missing electron from the full polarized  LL.  Thus it seems likely that the ground state at $f=4/5$ should be constructed out of skyrmions rather than of holes in the polarized LL. In particular, let us  consider a finite size skyrmion consisting of four missing electrons bound to three electrons in the reversed valley state.  If we add   such skyrmions to the $f=1$ state,  at a density of one skyrmion for every five flux quanta, we obtain a state with $f= 4/5$ and equal population of the two valley states. One possibility is that the skyrmions would form a Wigner crystal, in which case the Hall conductance would lie on the $\nu = -1$ plateau, and there would not be an FQH state at $f=4/5$. However, recent experiments do show clear evidence of an FQH state at $f=4/5$.  
Therefore, we believe that the FQH state at $f=4/5$  can be understood, roughly, as a Laughlin $m=5$ state formed from  hole-type skyrmions of the type  described   above.    

This state will be  fully spin-aligned but should be invariant under SU(2)  rotations in  valley space. Therefore, it cannot accommodate  pure valley skyrmions.  Any  skyrmions must  therefore involve the spin orientation, with an energy cost due to the Zeeman field.  Consequently, the state could have a substantial energy gap for charged excitations. 

Similar arguments suggest that  the ground state at $f=6/7$ should be a valley singlet  state, and that it might also be a quantized Hall state.  However, a quantized Hall state has not been observed experimentally at this fraction.

Unpublished calculations of finite systems by R. Morf \cite{morf13}    suggest that for pure Coulomb interactions, spin-singlet ground states should exist at $f=4/5$ and $f=6/7$, and that the probability density $P$ for two electrons to coincide is very small in both of these states, of order $10^{-3}$.  This implies that the effects of valley anisotropy should be negligible in these states, and it  is consistent with the skyrmion construction starting from $f=1$ state, as outlined above. 

\subsection{Other fractions with $0<f<1$}

At $f=4/9$, we see two candidate ground states.  One possible state would have equal numbers of electrons in each of two hyperspin states, occupying states in both valleys with spins parallel to the magnetic field. The ground state in this case could be described in the composite fermion picture as a state with two types of composite fermions occupying the lowest two effective LLs.   

An alternative state could have electrons in all four hyperspin states.  This state could be described by a generalization of the Halperin 332 state, in which the many-body wave function contains factors that vanish as $(z_i-z_j)^3$  for particles in the same hyperspin state, but vanish as $(z_i - z_j)^2$ for particles with different hyperspin states (here $z_i=x_i+iy_i$ is the complex coordinate of $i$th particle). Such a state would presumably minimize the Coulomb energy but would pay a price in Zeeman energy.  In principle, there could be a phase transition between the two types of states, if one varies the magnetic field at fixed  LL filling. An FQH state at $f=4/9$ has been observed in experiments, but no phase transition has been observed as yet. We also note that one could consider other states at $f=4/9$, e.g., ones in which three different hyperspin states are occupied. However, given the double degeneracy of CF levels at $f=4/9$, it seems unlikely that such states would be favorable. 

In a similar way, at $f=3/7$, we might consider a ground state with three different  hyperspins occupied, each with 1/7 electron per flux quantum,  or a state with all electrons in one hyperspin state, or a state with 2/7 electron per flux quantum in one hyperspin state, and 1/7 in another.  Similarly, at $f=3/5$, we could consider states with three different hyperspins occupied, with all electrons in one hyperspin state, or having two hyperspin states with filling factors of 2/5 and 1/5. 
All of these states would have charged skrymion excitations that cost no Zeeman energy, and little or no anisotropy energy.  Therefore, we might expect them to have only a small energy gap, and the FQH state would be easily suppressed by disorder or finite temperature. A FQH states at $f=3/7$ has not been observed yet, and there is only a faint hint of an incompressible state at $f=3/5$~\cite{Feldman13}.

\section{FQH States for $ 1 < f <2$}\label{fqhe2}

In the discussion below, we shall only consider candidate ground states  in which  only two fixed hyperspin states, $\sket {\chi_1}$ and $\sket{\chi_2}$ may  be occupied. (The remaining two hyperspin states are completely empty.)  States of this type are consistent with the symmetry of our Hamiltonian, and it seems like a reasonable assumption that the actual ground states will satisfy this restriction, but we do not have any rigorous argument that this must be the case at all filling factors.

Following our assumptions, the ground state at LL filling $f$ in the range $1<f < 2$  should be related by particle-hole conjugation to the ground state one would obtain for the same choice of $\sket {\chi_1}$ and $\sket{\chi_2}$
at  a LL filling $f ' = 2-f$, in the range $0<f'<1$.   In particular, the expectation value of the anisotropy energies  in the states at $f$ and $2-f$, for the given choice of hyperspin states,  would be related, following Eq. (\ref{ph2}), by
\be
\label{ean}
E^{\rm{an}}_{f} = E^{\rm{an}}_{2-f} + (f-1 )  E^{\rm{an}}_{2},
\ee
where $E^{\rm{an}}_{2}$  is the anisotropy energy of the  state at $f=2$ for this choice of  $\sket {\chi_1}$ and $\sket{\chi_2}$.
However, the conjugate state at $f'$ may not be the true ground state at that filling factor, because  the optimal choice of hyperspin states may be different at $f'$ than at $f$. 
We found previously, that the optimum  ground states for $0<f<1$  were obtained when the two constituent hyperspin states had opposite valley indices but both spins aligned with the magnetic field. However, if the Zeeman energy is small compared to the anisotropy terms at $f=2$, so that the AF state is the ground state at that filling factor, we may guess that for $1<f<2 $  we should choose the hyperspin states to be the same ones that are represented in the AF state, e.g.,  $K\uparrow$ and $K' \downarrow$, so the system is not fully spin-polarized.  (We assume, for the moment, that the Zeeman field is small compared to the anisotropy energy in the AF state, so we may ignore canting of the spins. We shall discuss effects of canting in a separate subsection below.)

In the following discussion we shall assume that  $\sket {\chi_1}$ and $\sket{\chi_2}$ are given by $K\uparrow$ and $K' \downarrow$, for $1<f< 2$.  With this choice, we see that 
$E^{\rm{an}}_{2} $ is the anisotropy energy of  the AF state at $f=2$, given by $E^{\rm{an}}_{2} =  - g_z / (2 \pi l_B^2 ) <0$.  

\subsection{$f=5/3 \,\,\,\, (\nu = -1/3)$}

The obvious candidate for this filling fraction is a maximally polarized  state $\sket{\Phi}$, where we align the spins quantization axis with the magnetic field, and  remove from the $f=2$ AF state one-third of the electrons in the hyperspin state with spin opposite to the magnetic field ($K' \downarrow$), while retaining the completely full LL of electrons in the hyperspin state $K \uparrow$ favored by the Zeeman interaction.  We may describe this state as a Laughlin state of hyperspin-aligned holes in the $f=2$ state.  The particle-hole conjugate of this state would be an $f=1/3$ state $\sket{\Phi '}$ where  all electrons are in the hyperspin state $K' \downarrow$.  That state will have no anisotropy energy, because of the Pauli exclusion principle, so the proposed  state  at $f= 5/3$ should have an anisotropy energy  $E^{\rm{an}}_{5/3} = (2/3)  E^{\rm{an}}_{2}$, according to (\ref{ean}).  

We argued, previously, that the true ground state at $f=1/3$ should have all electrons in a single  hyperspin state with spin parallel to the Zeeman field (here, spin $\uparrow$) and arbitrary choice for the valley pseudospin state.  If the Zeeman energy were absent, the state $\sket{\Phi ' }$  would  have the same energy as the true  ground state at $f=1/3$, so it must also be a state that  minimizes the combination of Coulomb and valley-anisotropy energies.  It follows by particle-hole symmetry, that at $f=5/3$  the wave function $\sket{\Phi}$ must also minimize these energies, among states within the class of we are considering, in which electrons are restricted to the hyperspin states  $K\uparrow$ and $K'\downarrow$.  As the state $\sket{\Phi}$ also minimizes the Zeeman energy, it is the correct ground state at $f=5/3$, at least within the class we are considering.

Although the ground state energy at $f=5/3$ is directly related to the energy of the ground state at $f=1/3$, the excitation energies will be quite different.  In particular,  a skyrmion at $f=5/3$ will inevitably engender a cost in Zeeman energy and/or anisotropy energy. Consequently, we expect that there will be a  substantial energy gap at $f=5/3$ and a well established quantized Hall state, in contrast with the situation at $f=1/3$. 

\subsection{$f=4/3 \,\,\,\,(\nu = -2/3)$  \,\, $(f=4/3)$   }

If we restrict our considerations to many-body states with only two occupied hyperspin states, the ground state at $f=4/3$ should be related by particle-hole conjugation to the ground state at $f=2/3$ for the same choice of hyperspin states.  However, the optimal choice of the hyperspin states should be different in the two cases.  We found previously that the optimum  ground state at $f=2/3$ was obtained when the two constituent hyperspin states had opposite valley indices but both spins aligned with the magnetic field. However, at $f=4/3$, if the Zeeman energy is small compared to the anisotropy terms, and the AF state is the ground state at $f=2$, we may guess that it would be better to choose the hyperspin states to be the ones that are represented in the AF state, that is $K\uparrow$ and $K'\downarrow$ (we will justify this below, in the subsection on spin canting).  Now, if we choose the spin quantization axis to be parallel to the Zeeman field, the state with fully polarized holes can  have a lower Zeeman energy  than the state with  equal populations of the two hyperspin states.  Thus the Zeeman field can now drive a phase transition between the two states. 

To make this more precise, we must again consider the anisotropy energy, which will be non-zero in both of the states we consider.  From Eq.(\ref{ph2}), we see that the anisotropy energy of  a state at $f=4/3$ is related to that of the  conjugate at $f=2/3$ by
\be
E^{\rm{an}}_{4/3} = E^{\rm{an}}_{2/3} + (1/3)  E^{\rm{an}}_{2}.
\ee
Since there can be no anisotropy energy   in the fully polarized state at $f=2/3$, we see that  anisotropy energy for the maximally polarized  state at $f=4/3$ for the same value of the magnetic field, will be equal to one-third of the anisotropy energy at $f=2$.   The anisotropy energy for an $f=2/3$ state with equal occupations of $K\uparrow$ and $K'\downarrow$ would be given by 
$E^{\rm{an}}_{2/3} = g_z  P_{2/3}  = P_{2/3}   E^{\rm{an}}_{2}$, where $P_{2/3}$ is the probability  density for finding two electrons at the same position in space in the $f=2/3$ state. The value of $P_{2/3}$ does not depend on the choice of hyperspin states, as the orbital part of the wave function is determined by the SU(4) invariant part of the Hamiltonian, and we  have previously argued that the value of $P_{2/3}$ should be very small. 
Consequently, we expect that the {\em difference} in the anisotropy energies of the maximally polarized state and the unpolarized state at $f=4/3$, given by $| P_{2/3}  E^{\rm{an}}_{2}|$,  will be small compared to the energy difference due to the Zeeman field.  Thus the total energy difference between the fully polarized state and the pseudospin-balanced state is given by the Zeeman energy, which is $\propto B$ and favors the former, and the Coulomb energy, $\propto B^{1/2}$, which favors the latter.  Consequently, there can be a transition between the two states, with the polarized state favored at larger magnetic fields.  

In neither of these states can there be  skyrmions with negligble cost in anisotropy energy or Zeeman energy.  Therefore, both states should have a  reasonable energy gap, and should lead to well established quantized Hall states at low temperatures.

\subsection{$ f=8/5 \,\,\,\, (\nu = -2/5)$}

The situation at $\nu=-2/5$ should be very similar to that at $\nu=-2/3$. Again, we have two competing states: a balanced state where there are equal numbers of holes in $K\uparrow$ and $K'\downarrow$, and a maximally polarized state with all the holes in one of the two states.  As there will be essentially no probability for two holes to be at the same point of space in either of these states, there will be no difference in their valley-anisotropy energy. As at $\nu = -2/3$, the maximally polarized state will be favored by the  Zeeman energy, while the balanced state should be favored by  the Coulomb energy.  Again, there can  be a transition between  the two states as one varies the magnetic field at fixed filling factor.

\subsection{Other fractions}

Quantized Hall states should be expected at many odd-denominator  ``Jain fractions"  in the range $-1/3 > \nu > -2/3$ according to the composite fermion picture. In particular, if one considers the integer QH state at $f=2$ as the reference state, and attaches two Chern-Simons flux quanta to each hole inserted in that state,  one expects to find quantized Hall states at fillings $\nu = - p/(2p \pm 1)$, where $p$ is an integer which describes the total number of holes per quantum of flux in  the effective magnetic field $B_{\rm{eff}}$ = $B / (1 \pm  2p)$.  If $p>1$, the  holes can be distributed in various ways between the  $K\uparrow$ and $K'\downarrow$ states.  As we found for $\nu = -2/5$ and $\nu=2/3$, which correspond to $p=2$ with the two choices of $\pm 1$, we expect that states with different polarizations will differ very little in their anisotropy energies, but that states with greater imbalance  will be favored by the Zeeman field whereas states with lesser imbalance  will be favored by the Coulomb energy. Thus  we would expect a series of phase transition at all these filling fractions,  as discussed  in Ref.~[\onlinecite{Feldman13}].  

A value of $\nu$ that  is intermediate between two Jain fractions can be described by means of a fractional value of  $p$. As discussed in \cite{Feldman13} the ground state of such a system can be described in the composite fermion picture as  state where an integer number of composite fermion LLs are filled for one of the two hyperspin species while there is a fractional occupation of the highest LL for the other species.  When $B$ is varied, phase transitions occur as the fractional occupation  shifts from one species to the other. 
It is also worth noting that oscillations that may be interpreted as signatures of such phase transitions were observed\cite{Feldman13} even in region very close to $\nu = -1/2$ where quantized Hall states were not apparent. 

\subsection{Effects of spin canting}

In the previous discussion, we assumed that the FQH states with equal population of $K,K'$ valleys are AF. Now we consider the effect of Zeeman interaction on such valley-balanced states, and show that it only leads to weak canting of spins in the direction parallel to magnetic field. This canting changes the energy of the state very slightly and does not influence the phase transitions between valley-balanced and valley-imbalanced FQH states in the interval $-2/3<\nu<-1/3$. In this Subsection, we also discuss possible candidate states at filling factors $-1<\nu<-2/3$. 

In order to consider the possibility of spin canting, in the regime $1<f<2$, we must consider states in which the spins on the two sublattices can have orientations which are no longer  opposite to each other.  We shall assume the sublattice spins and the magnetic field direction to be coplanar, and we shall denote by $\theta_K$ and $\theta_{K'}$ the angles between the spins and the magnetic field, so that the angle between the spin orientations is $\theta_{K'}-\theta_K$.
We define $f_K$ and $f_{K'}$ to be the filling factors on the two valleys, which, without loss of generality, we may restrict to satisfy $0< f_{K'} \leq f_{K} \leq 1 $, and $f_{K} + f_{K'} = f$.  

 We wish to understand how various terms in the energy depend on the the angles $\theta_{K}$ and $\theta_{K'}$. For fixed values of $f_{K}$ and $f_{K'}$, the SU(4) Coulomb energy will be independent of these angles. The Zeeman energy, per flux quantum, will be given by
\be
{\cal{E}}_Z= - E_Z   ( f_K   \cos \theta_K + f_{K'}  \cos \theta_{K'})      
\ee
To evaluate the energy due to the anisotropic valley interaction, we may make use of Eq. (\ref{ean}).  The three terms in this equation should all be evaluated for the same angles  $\theta_K$ and $\theta_{K'}$. The term with total filling $2-f$ should have sublattice fillings $(1-f_A)$ and $(1-f_B)$ while the term at $f=2$ must have both hyperspin states filled. We have previously  observed that the anisotropy energy for FQH states in the range  $0<f \leq 2/3$ should be extremely small.  Therefore, we may neglect the term with total filling $2-f$ in Eq.(\ref{ean}) and write,  for $4/3 \leq f \leq 2$,
\be
\label{eanc}
E^{\rm{an}}_{f}(\theta_K,\theta_{K'}) =  (f-1 )  E^{\rm{an}}_{2} (\theta_K, \theta_{K'}) \, .
\ee
Since valley-anisotropic interaction  is invariant under a uniform rotation of all electron spins,  the anisotropy energies can only depend on $\theta_{K'} - \theta_K$, the angle between the spins.
Moreover, we may write~\cite{Kharitonov11}
\be
E^{\rm{an}}_{2} (\theta_{K'}-\theta_K) =- (2 \pi l_B^2 )^{-1} \left( g_z  + g_{\perp}  [1 +\cos (\theta_{K'} - \theta_K)] \right) \, .
\ee
As we have assumed $g_{\perp} < 0$,  this energy is clearly minimized when $\theta_{K'}-\theta_K = \pi$, in which case we have
\be
E^{\rm{an}}_{f} = -  (2 \pi l_B^2)^{-1} (f-1)  g_z   \equiv  E^{\rm{an}}_{f} (\pi)   \, .
\ee

 If  $f_K=f_{K'}$,  and $E_Z$ is small but nonzero, 
 the total energy will be minimized by a choice $\theta_{K'}= -\theta_K = ( \pi/2 - \alpha)$ where the canting angle  $\alpha$ is proportional to the quantity $\gamma$ defined in Eq.(\ref{eq:gamma}). More precisely, one finds
  \be
  \label{sina}
  \sin \alpha = \cos \theta_K = \frac {\gamma f}{4 (f-1)}\, ,
  \ee
  provided that the right hand side is $\leq 1$.  If the right hand side of (\ref{sina}) is greater than 1, one finds that $\theta_K=\theta_{K'} = 0$; both spins are aligned with the magnetic field and the state is no longer canted.
 The sum of the Zeeman and anisotropy energies for the canted state may be written as
  \be
E^{\rm{an}} + {\cal{E}}_Z =  E^{\rm{an}}_{f} (\pi) - \frac{E_Z f }{2} \sin\alpha \, ,
\ee
so the decrease in energy due to the canting will be proportional to $\gamma E_Z$, when $\alpha$ is small. Since $\gamma\approx 0.1$, the energy gain from canting is indeed quite small for any valley-balanced state ($f_K=f_{K'}$) in the range $1<f<4/3$. 
  
If $f_K \neq f_{K'}$ and if $\gamma$ is sufficiently small, one finds that the  the lowest energy state will not have canted spins but rather will have collinear spins with  $\theta_A=0$ and $ \theta_B = \pi$ (as we assumed in previous Subsection).  In this case,  spins of electrons in the $K$ valley are aligned with the magnetic field, and spins in the $K'$ valley point in the opposite direction, as we have assumed in our previous discussions.  
The sum of the Zeeman and anisotropy energies for the collinear state will have the form
\be
\label{uncant}
E^{\rm{an}} + {\cal{E}}_Z =  E^{\rm{an}}_{f} (\pi) - E_Z (f_K - f_{K'}) \, .
\ee
Since the energy gain due to Zeeman term is linear in $E_Z$, this state will have a lower energy than the canted state, for sufficiently small values of $\gamma$. 
We find that the collinear state, with $\theta_K=0$ and $\theta_{K'}= \pi$  is stable  if
\be
f_K  - f_{K'} >  \gamma  \frac {f_K f_{K'}} {f-1} . 
\ee
For  a fixed value of $f$ and $\gamma$, this inequality will be satisfied provided that $f_K-f_{K'}$ is larger than a critical value
 $\delta f_c $,  which, for $\gamma = 0.1$ and  $4/3 < f <2$  will be well approximated by $ \delta f_c  \sim  \gamma f^2 / (4f-4)$. For such states, the phase transitions between FQH states with $f_K=f_{K'}$ and $f_K\neq f_{K'}$ will occur as a function of magnetic field, very similar to our discussion above. 
 
For  $f_K-f_{K'} < \delta f_c$, the ground state for fixed $f_K$, $f_{K'}$ will have spins that are canted from each other, and the ground state  energy will be slightly lower than it would be for the uncanted state. However the qualitative behavior of the energy as a function of $f_K-f_{K'}$  will not be altered from that of (\ref{uncant}): the sum of the Zeeman and anisotropy energies will still decrease monotonically as  $f_A-f_B$  is increased.  Thus, in this case we would also expect to see the same series of transitions as a function of magnetic field, at fixed filling factor, with at most small changes in the locations of the transitions when one of the states involved  has $f_A-f_B < \delta f_c$.   

We note that for $\gamma =0.1$, we find $\delta f_c < 0.3$  for all $f$ in the range $4/3<f<2$, and $\delta f_c \to 1$ for $f \to 2$.  Thus at $f=9/5$ and $f=5/3$, where one expects the ground state to have all holes in one valley, so that $f_K=1$ and $ f_{K'}= f-1$, we have $\delta f > \delta f_c$, and the majority spins will be fully aligned with the magnetic field, as we have assumed in the previous sections.  This will also be true for the unbalanced states at $f= 12/7$ and $f=8/5$, where all holes  sit in one valley.  The balanced states at these filling fractions will be antiferromagnets with spins that are only slightly canted,  which is again consistent with our previous assumptions.

Now we turn to discussion of filling factor interval $1<f<4/3$. First, we note that the above picture might change completely if one considers filling factors where $f$ is very close to 1. In this case the right-hand side of (\ref{sina}) may exceed unity, so for $f_K=f_{K'}$,  the canted state will be replaced by a ferromagnetic state, where the spins are fully aligned with the magnetic field, and the valley state is a singlet. For the spin-aligned state, one finds
\be
\label{aligned)}
E^{\rm{an}} + {\cal{E}}_Z =  E^{\rm{an}}_{f} (\pi) - 2 (f-1) g_{\perp} ( 2 \pi l_B^2)^{-1}    - f E_Z .
\ee
If $\gamma = 0.1$, however, the state with $f_K=f_{K'}$ will be fully polarized only for $f-1 \leq 1/39$, and the canting angle will be relatively small $(\cos\theta_K \leq 1/2)$ for filling factors as close to 1 as $f-1 \geq 1/19$. A state with maximum valley polarization, i.e., $f_K=1$ and  $f_{K'} = f-1$ should have a fully aligned  spins, for any  value of $f-1$. We have previously noted that for typical experimental parameters, the lowest energy electron-like excitations should be skyrmons with a small number of electrons in the minority valley and a small number of overturned spins. Therefore, it seems likely that the ground  state at  a very small positive value of $f-1$ will be a state which is partially valley polarized and partially spin polarized. However, the ground state under these conditions  would be a Wigner crystal rather than an FQH state.  
 
At filling factor  $f=6/5$, we  expect that there should  be a  valley-balanced ground state with $f_K=f_{K'}$,  whose  spatial wave function would be the particle-hole conjugate of the valley-singlet state at $f=4/5$.  However, we expect that the occupied hyperspin states at $f=6/5$ would be different, leading to a weakly canted AF state rather than a spin aligned state (as for $f=4/5$).  In principle, for large values of $B$, there could be a series of transitions to states with $f_K-f_{K'} \neq 0$.  However, it is likely that the Coulomb cost for such states will be high, and we would not expect to see such states for reasonable values of the magnetic field. At present no experimental observation of an  FQH state of any type has been reported  at $f=6/5$.  

We note that the parameter $\gamma$ will be increased if one applies a magnetic field parallel to the graphene sheet that is large compared to the perpendicular field. For sufficiently large values of $\gamma$ we expect that in  all states with $\nu \leq 0 $, the electrons in the zeroth Landau level will all have spins completely aligned with the magnetic field.  In this case, one would expect that particle hole symmetry is restored between  states $f$ and $2-f$.


\section{Disorder effects}\label{disorder}

The energy gap  $E_g$ of a quantized Hall state may be defined as the energy cost to add a quasiparticle and a quasihole very far apart.  In the limit of vanishing disorder, the energy gap may be obtained from incompressibility or transport measurements.  In transport measurements, the low temperature behavior of the longitudinal conductivity $\sigma_{xx}$ is thermally activated, varying as $e^{-E_g/ 2k_B T}$.  Measurements of the compressibility using an SET~\cite{Feldman12,Feldman13}, should show a discontinuity $\Delta \mu$ in the electronic chemical potential equal to $E_g \, e/e^*$ at the quantized Hall density, where $e^*/e$ is the charge of the quasiparticle in units of the electron charge. (For all cases considered here,  $e/e^*$ is equal to  the denominator of the filling fraction $f$.)   

The effects of disorder on various types of measurements are only partially understood, even in the much studied case of GaAs systems. In principle, any amount of disorder should lead to a finite density of localized states in the energy gap, so that $d \mu /dn$ should be finite even when $n$ coincides with an ideal  quantized Hall value, and the associated  jump in $\mu$ should occur over a finite range of densities, inversely proportional to  $d \mu /dn$. Moreover, the peaks and valleys in $\mu$ on the two sides of the quantized Hall density, which would be sharp cusps in the ideal system,  are typically rounded over a  similar finite density range. 
In an SET measurement, the width of this range can be minimized, and the slope $d \mu /dn$, can be maximized, by placing the SET at a location that is relatively free of disorder.   

The behavior of $d \mu /dn$ described above can be seen in panel (a) of Fig.1 below, which shows data for a suspended graphene sample near $\nu = 0$, at three values of $B$.   Qualitatively similar behavior is seen in SET measurements  near other quantized Hall states in graphene, as well as in GaAs.  However, we do not have a good quantitive understanding of the density of states in the gap or of the extent of rounding in any of these cases.

The effects of disorder on transport measurements are even more difficult to interpret. For an integer quantized Hall effect, in a system of non-interacting electrons, one might expect to see activated behavior at low temperatures,  determined by a  mobility gap, i.e., the energy difference  between extended states in the filled and empty LLs.  However, in an interacting electron system, or in the presence of phonons, one would expect that the conductivity will be described by variable range hopping at the lowest  temperatures. Activated behavior is typically seen in experiments over an intermediate temperature range,  typically about a factor of two in temperature, but the interpretation of the activation energy, and its relation to the energy gap of an ideal system are not  at all clear.~\cite{dambrumenil}

Disorder effects in graphene may be particularly important, and may present special features, in the cases where the low energy charged excitations are large-area valley skyrmions. 
Below, we shall explore the effects of disorder on quantum Hall states in graphene at different filling factors (both integer and fractional). 

The spin- and valley-polarized states that are predicted for various filling fractions in the range   $0<f\leq 1$ imply  spontaneous breaking of  a continuous $SU(2)$ valley symmetry. As a result, these states will have low energy Goldstone modes, and they will be very sensitive to valley-selective disorder potential. By the Imry-Ma argument~\cite{ImryMa},  arbitrarily weak disorder will  destroy the long-range valley order, and a ground state with spatially varying valley polarization should form. The precise nature of the textured ground state, and the structure of low-energy excitations depend on the dominant type of the valley-selective disorder. However, generally such textured ground states will have low-energy charged excitations. This can significantly reduce the robustness of a quantum Hall state, or even completely destroy it. 

In contrast, valley-singlet states at $0<f<1$ ($f=2/5,2/3,4/9$) do not have a spontaneously broken symmetry.  Therefore they should be fairly robust: weak valley-selective disorder might slightly reduce the value of the energy gap, and might introduce isolated localized states in the gap,  but should not  change the nature of the ground state. 

In the range $1<f\leq 2$, we find two types of quantized Hall states; both types turn out to be insensitive to weak disorder. First, states with unequal populations of the two valleys have a spontaneously broken valley symmetry, but the anisotropy energy reduces this from a continuous  SU(2) symmetry to a discrete Z(2) symmetry.  The electron spins  in the majority and minority valley states are aligned, respectively,  parallel and antiparallel to the external magnetic field, so there will be no Goldstone mode associated with either the spin or  valley degree of freedom.   For states with a spontaneously broken Z(2) symmetry, the Imry-Ma argument suggests that disorder is marginal in two dimensions. Although domains should be induced, in principle, by any amount of disorder that couples to the difference in populations of valleys $K$ and $K'$, the size of such domains should be exponentially large when the disorder is weak, so domain walls should  be few and far apart~\cite{Imbrie}. 

Second, for states with equal population of valleys $K$ and $K'$, we expect that the ground state will be a canted antiferromagnetic state, where the spins have a  spontaneously broken U(1)  symmetry in the plane perpendicular to the magnetic field.  This broken symmetry gives rise to a  gapless mode in the spin sector, but the perpendicular spin component would not couple to disorder produced by adatoms, impurities or strains.  (In addition, even if one produces a skyrmion in 
antiferromagnetic spin order  by forcing the direction of antiferromagnetic  order to wrap around  the unit sphere, such a skyrmion would not carry an electric charge.) The broken valley symmetry in an AF state is again pinned by the valley anisotropy terms, so that only a Z(2) symmetry is spontaneously broken. Thus there are no Goldstone modes in the valley sector, so there would  be  no  charged skyrmions with large area and low energy.  Moreover, since the $K$ and $K'$ electron populations are equal, the broken Z(2) symmetry should not couple to potential fluctuations that break the valley symmetry. 

Because of this, all fractional states in the interval $1 < f \leq 2$,  both valley-unpolarized and partially valley-polarized states, should be robust in the presence of weak disorder. For the case of valley-polarized states,  spatially varying configurations of the order parameter can be created only if  the disorder becomes stronger than valley anisotropy, and the valley-unpolarized states  should be even more robust.   

We focus now on the valley polarized states with filling fraction $0<f\leq 1$, which are most susceptible to valley-selective disorder.  It is convenient to specify the orientation of the  valley polarization by a unit vector $\vec{n}(\vec{r})$ in pseudospin space, whose components describe the normalized expectation value of   the operators   $\hat{\psi}^\dagger (\vec{r}) \hat{\tau}_i \hat{\psi} (\vec{r})$, for $i = 1,2,3$. 
 We distinguish two types of valley-selective disorder: (i)  an in-plane random field that  couples to $n_x$ and/or $n_y$, and therefore mixes valleys $K,K'$, and (ii) a random field that distinguishes the two valleys and therefore couples to  $n_z$. The first type of disorder could arise from adatoms situated at the center of hexagons or on hexagon sides in graphene lattice (e.g., various metallic adatoms). The second type of disorder may stem from adatoms that are situated on the sites of graphene lattice (e.g., H-adatoms), or from vacancies. Alternatively, such disorder can arise due to random short-range strain~\cite{Abanin07}; such strains are present in graphene on SiO$_2$ substrates, but probably are less important in suspended graphene. Due to the different origin of two disorder types, one or the other will likely prevail in any experimental situation.  

\begin{widetext}

Assuming that the spin degree of freedom is frozen by the Zeeman interaction, we can write an effective energy functional for the $SU(2)$ valley order parameter ${\bf n}({\bf r})$: 
\be\label{eq:eff_H}
H=\frac{\rho_s}{2} \int d{\bf r} [\nabla {\bf n}({\bf r})]^2   +f\int d{\bf r} {\vec h}_{\perp}({\bf r}) {\vec n_{\perp}}({\bf r})+f\int d{\bf r} h_z({\bf r}) n_z({\bf r}) \, ,
\ee 
where  ${\vec n_{\perp}}=(n_x,n_y)$.
We have introduced an explicit  factor $f$  in front of the random potential terms to emphasize that the coupling to the order parameter for a state that is fully valley polarized should be proportional to $f$, for a fixed value of $B$.   

\end{widetext}

We shall assume that the random fields ${\vec h_{\perp}}$, $h_z$, have short-range correlations on the scale of magnetic length. This is obvious for the case when disorder is due to adatoms, and it may be correct in many cases for local strains produced by inhomogeneities in the coupling to a substrate.   Thus we may write
\be\label{eq:h_p}
\la h_\alpha({\bf r}) \ra=0, \,\,\,\la h_\alpha({\bf r}) h_\alpha ({\bf r'})\ra =V_{0}^2 n_{{\rm imp},\perp} \delta({\bf r}-{\bf r'}), \,\, \alpha=x,y. 
\ee
\be\label{eq:h_z}
\la h_z({\bf r}) \ra=0, \,\,\,\la h_z({\bf r}) h_z ({\bf r'})\ra =V_{0}^2 n_{{\rm imp},z} \delta({\bf r}-{\bf r'}), 
\ee
where  $n_{{\rm imp},z}$  and  $n_{{\rm imp},\perp}$ may be interpreted as  densities of the two types of impurities, and $V_0$ represents the root-mean-square value of the coupling to a single impurity. The quantity $V_0$ has dimensions of an energy, and it will be proportional to the density  of magnetic flux, $B/\Phi_0$.  Strictly speaking,  $V_0$ should be different for different types of impurities, but we may ignore this distinction without consequence for  the following discussion. 

Following the Imry-Ma argument, the disorder potential will destroy  long-range valley order by creating domains. The domain configuration is determined by the disorder realization, and the structure of domains depends on the dominant type of disorder. 
The typical domain size $\xi$ can be estimated by noting that the energy gained due to the adjustment of the order parameter to the local disorder configuration, $E_{\rm dis}\sim  f V_0 \xi \sqrt{n_{\rm imp}}$, should be comparable to the stiffness energy $\sim \rho_s$. (Notice that because of $SU(2)$ symmetry, the order parameter varies on the scale $\xi$.) Thus,  the typical domain size, given by
\be\label{eq:domain_size}
\xi\sim \frac{\rho_s}{f V_0} \frac{1}{\sqrt{n_{\rm imp}}}, 
\ee
is proportional to the valley-stiffness $\rho_s$, and inversely proportional to the filling fraction. Here $n_{\rm imp}={\rm max}(n_{{\rm imp},z}, n_{{\rm imp},\perp})$. 

Let us consider two limiting cases: (1) $n_{\rm imp,\perp}\gg n_{{\rm imp},z}$, that is, in-plane disorder dominates, and (2) $n_{\rm imp,\perp}\ll n_{{\rm imp},z}$, the random $h_z$  field dominates. The two cases are characterized by different symmetry and different structure of the textured ground state, as discussed below. However, in both cases there are low-energy excitations. 

To gain some intuition about case (1) (dominant in-plane disorder), let us first completely neglect $h_z$. In this case, the order parameter in different domains will be pointing predominantly in-plane. This will lead to the formation of vortices, where the order parameter rotates by $2\pi$. At the vortex cores, the order parameter must tilt out of the $xy$ plane. Given the $n_z \to -n_z$ symmetry of the Hamiltonian, there are two degenerate configurations of the order parameter for a single vortex ($n_z$ at the vortex core can be either positive or negative). Such configurations correspond to {\it oppositely} charged merons with charge $\pm e/2$. This has two implications. First, 
the textured ground state has some density of frozen charges of each sign. Second,  for an isolated vortex, changing the charge from positive to negative sign costs no stiffness energy. It should be noted, however, that the precise configuration of the order parameter will be determined by the balance between stiffness energy, disorder, and long-range Coulomb interaction. 

A weak random $h_z$ field will pin the direction of the $n_z$ component in the meron cores in accordance with the local disorder configuration. As a result, the meron and anti-meron state will no longer be degenerate, but rather, will be separated by a small energy difference. The precise value of this energy difference  will depend on the local disorder and on the Coulomb interaction with nearby frozen charges. Due to randomness, the energy difference  will be distributed in some energy interval $(-E_{sk}; E_{sk})$; almost certainly, $E_{sk}$ will be much smaller than the skyrmion energy in a clean system, $E_{sk}^0=4\pi \rho_s$. As we discuss below, this may strongly enhance the compressibility of certain fractional states, such that they will appear to be fragile or absent in experiments.

Let us define a net skyrmion density $N_{sk}({\bf r})$  as the difference in the number of negative and positive merons, per unit area. 
If we assume that the  energy differences for converting a positively charged skyrmion  to a negative one are uniformly distributed in the  interval  $(-E_{sk}; E_{sk})$, we  find that the energy cost to produce a net imbalance in the density of positive and negative merons should be given by 
\be\label{eq:skyrmions_functional}
E[N_{sk}]=\frac{K}{2}\int N_{sk}^2({\bf r}) d{\bf r},  
\ee
where  $K$ can be estimated in terms of the typical energy of a single extra skyrmion and the typical size of a single domain, 
$$
K\sim E_{sk}\xi^2.
$$
  The chemical potential $\mu$ for electrons is defined as the derivative of the total energy  with respect to electron number, after subtracting any contribution from the interaction of the electrons  with a  macroscopic electrostatic  potential.  If we assume that  fluctuations in the electron density  arise purely from fluctuations in the skyrmion density, and are given by $\delta n = (e^*/e) N_{sk}$,  then we obtain an inverse compressibility given by
   \be\label{eq:compressibility}
\frac{d\mu}{dn}\sim \frac{e^2}{{e^*}^2}{E_{sk}} \xi^2. 
\ee
Taking into account Eq.(\ref{eq:domain_size}), we rewrite the above equation as
\be\label{eq:comp2}
\frac{d\mu}{dn}\sim \frac{e^2}{V_0^2 n_{\rm imp}}\times \frac{\rho_s^2 E_{sk}}{{f^2 e^*}^2}.
\ee
Noting that $E_{sk}=C\rho_s$, where $C$ is a small numerical constant, we obtain the final form of inverse compressibility
\be\label{eq:comp3}
\frac{d\mu}{dn}\sim \frac{Ce^2}{V_0^2 n_{\rm imp}}\times \frac{\rho_s^3}{{f^2 e^*}^2}.
\ee

This simple result allows us to make predictions regarding the sensitivity of different ferromagnetic states at $0<f\leq 1$ to disorder. For example, let us compare the integer state $f=1$ (which is observed experimentally in compressibility measurements) to the fractional state $f=1/3$ (which is absent). The $f=1$ is characterized by valley stiffness $\rho_s=e^2/16\sqrt{2\pi}l_B$, while $f=1/3$ state has a stiffness which  is approximately 12 times smaller~\cite{Wojs02,Toke06} (note however that this number should be viewed only as an estimate, due to large finite-size effects in numerical calculations~\cite{Wojs02,Toke06}). Therefore, the ratio of the inverse compressibility for these two states is given by
\be\label{eq:ratio}
\frac{(d\mu/dn)|_{f=1}}{(d\mu/dn)|_{f=1/3}}\approx \frac{12^3}{81}\approx 21. 
\ee
Thus, in the disordered case the skyrmion contribution to compressibility is much smaller for the integer state $f=1$; this is due to the large valley stiffness of this state, which gives rise to large 
domains and smaller density of frozen charges and low-energy excitations. This explains why the disorder-induced low-energy skyrmions may suppress the $f=1/3$ state in experiment, but not the $f=1$ state.

Turning, now,  to the case (2) above, where $h_z$ is the dominant random anisotropy term, we first note that neglecting in-plane disorder completely, the Hamiltonian (\ref{eq:eff_H}) is characterized by a $U(1)$ symmetry with respect to rotations of the order parameter in the $xy$ plane. This symmetry reflects the fact that the energy does not change if we globally change the relative phase of $K$ and $K'$ states; such that if $\alpha ({\bf r}) |K\ra +\beta ({\bf r}) |K'\ra$ is a ground-state configuration for a given order parameter, then $\alpha ({\bf r}) |K\ra + e^{i\phi}\beta ({\bf r}) |K'\ra$ is also a ground state. 
An important difference compared to case (1) is that in the ground state there are no textures and charge excitations: the order parameter rotates only along some meridian of the Bloch sphere, not enclosing any area (such that topological index of the order parameter configuration is zero). However, the non-uniform configuration of the order parameter in the ground state will be favorable for low-energy meron-anti-meron pairs, where phase $\phi$ rotates in different directions in neighboring domains. Therefore, low-energy excitations are expected in this limit as well.  

It is also worth noting that if small $n_{\rm imp,\perp}$ is added,  larger  super-domains  will be formed in which the order parameter has a preferred component in the $x$-$y$ plane.    Such super-domains will enclose many domains with preferred $n_z$ orientation. As a result, non-zero density of charges will be introduced in the ground state. It is not obvious whether the lowest-energy excitations will be textures in "small" domains, the size of which is set by $n_{{\rm imp},z}$, or the merons/anti-merons forming in "large" domains, the size of which is set by $n_{\rm imp,\perp}$. In either case, we expect that these excitations will have much lower energy than $E_{sk}^0$. 


The discussion above suggests that in the limiting cases when one kind of disorder dominates, there are low-energy charged excitations. In the intermediate regime, the textured ground state will have a certain density of charges (on the order of an electron charge per domain). Although it is difficult to make any quantitative prediction, it is reasonable to expect that the energy cost for creating skyrmions will be greatly reduced as well in this case.  

At this point, we are unable to say whether the enhanced compressibility caused by skyrmions in the presence of disorder is strong enough to account for the strong suppression of the  FQH states  with odd-numerator filling fractions in the range $0<f<1$  in the SET measurements reported in Refs.~[\onlinecite{Feldman12,Feldman13}].  We also have little information about the
mobility of these skyrmions, so it is difficult to estimate their contribution in  transport measurements. Nevertheless, the apparent  absence of an FQH signature at $f=1/3$ in transport measurements is striking, and it seems likely that  this absence is also due to the presence of low energy skyrmions at this value of $f$. 

It is also worth noting that suppression by disorder of  quantum Hall states with a spontaneously broken symmetry has been observed in other multi-valley systems~\cite{Shayegan1,Shayegan2,Si1,Si2}. For example, in AlAs-2D electron system, which is characterized by a double valley degeneracy, the measured activation energy gaps of certain fractional states in valley-symmetric case can be reduced by a factor of 15 compared to the case when valley symmetry is broken by strain~\cite{Shayegan2}. It should be noted, however, that the quantum Hall ferromagnet in AlAs has a lower, $Z_2$ symmetry compared to the quantum Hall ferromagnet in graphene at $0<f\leq 1$, with the easy-valley anisotropy being generated by the asymmetry of the band structure~\cite{Abanin10}. This may change the structure of domains, domain walls and their excitations, compared to the case of graphene.  
We also note that experiments on Si (111)~\cite{Si1} and Si (100)~\cite{Si2}
have observed FQH states at various fractions, almost entirely with even
numerators.  However, a strong FHE state has been reported at nu = 1/3 in
at least one experiment on Si (100)~\cite{Si3}.

\section{Experimental measurement of energy gaps at $\nu=0$ and $\nu=-1$.}\label{skyrmions}

Below, we present experimental measurements of the energy gaps of the broken-symmetry states at filling factors $\nu = 0$ and $-1$ in suspended graphene.  We extract these energy gaps using a scanning SET.  Details of the device fabrication and the measurement technique are provided in Ref. [\onlinecite{Feldman12}].  Briefly, we use a back gate to modulate the carrier density in the graphene, and monitor the resulting change in current through the SET to measure the local chemical potential $\mu$. This is accomplished using a DC feedback technique to maintain constant current through the SET; the change in sample voltage required for this feedback provides a direct measure of the chemical potential of the graphene.

Figure 1a shows the chemical potential as a function of carrier density at several representative magnetic fields.  As the chemical potential approaches $\nu = 0$, it first decreases, then abruptly rises, and finally declines again before stabilizing.  The regions of decreasing chemical potential (negative compressibility) are understood~\cite{F2} to arise from electron-electron interactions, which become especially pronounced at the low electron and hole densities surrounding $\nu = 0$.  Similar behavior occurs for $\nu = -1$ (data not shown).  The energy gaps at $\nu = 0$ and $-1$ are defined as the difference between the minimum and maximum chemical potential around each incompressible state, and they are plotted as a function of magnetic field in Figs. 1b and c, respectively.

The magnitudes of the energy gaps are significantly larger than has been previously reported~\cite{Young12}, which reflect both the high sample quality and also the low dielectric environment of suspended graphene.  At $B=12\, {\rm T}$ the gap at $\nu=-1$ is comparable to the theoretical skyrmion gap $\Delta_{\nu=-1}=\frac{\Delta_0}{2}\approx 23\, {\rm meV}$, estimated for dielectric constant $\epsilon=5.24$. Energy gaps at both filling factors are well-modeled by $B^{1/2}$ scaling (red fits), consistent with expectations for interaction-driven states.  Both fits, however, require subtraction of a constant quantity, giving  negative intercepts of a few meV.  A negative intercept in the measured energy gap may arise in part from disorder broadening of the cusps in chemical potential~\cite{F4}, although we note that the $\nu = 0 $ offset of -11.1 meV implies significantly larger disorder than was estimated for other samples with comparable quality~\cite{Dean11}.  

We would like to attribute the difference in energy gaps between  $\nu=0$ and $\nu=-1$ to the effects of skyrmions, which occur in the latter case.  According to mean field theory, neglecting the effects of valley anisotropy and Zeeman energy, the energy gaps for creating a single  electron or hole should be equal at these filling factors.  However, as discussed in the text, the lowest energy charged excitations at $\nu=-1$ should be large area valley skyrmions, which lead to an energy gap that is one-half the single-electron gap.  At $\nu=0$, we expect that large-area skyrmions will be suppressed due to the effects of valley anisotropy. The fact that the observed energy gaps at $\nu=0$ are more that twice the energy gaps at $\nu=-1$ may well be due to effects of residual valley-selective disorder, which would affect the skyrmions at $\nu=1$ most strongly, as discussed above.

\begin{widetext}

\begin{figure}[t]
\begin{center}
\includegraphics[width=6in]{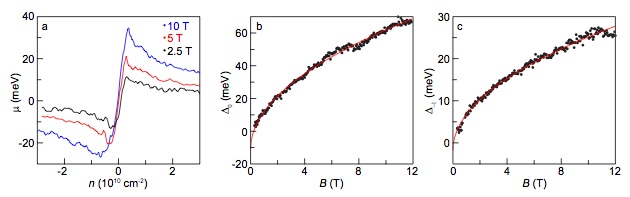}
\vspace{-0.7cm}
\caption{(a) Chemical potential $\mu$ as a function of carrier density $n$ at magnetic fields $B$ = 2.5, 5 and 10 T.  The energy gap is given by the difference between minimum and maximum chemical potential.  (b) Step in chemical $\Delta \mu$ at filling factor $\nu = 0$ as a function of magnetic field.  (c) Step in chemical at filling factor $\nu =- 1$ as a function of magnetic field.  The red curves in panels (b) and (c) are fits with $B^{1/2}$ scaling: $\Delta \mu =\left( 23.1B^{1/2}-11.1\right)$ meV, for $\nu= 0$,  and $\Delta \mu =\left( 8.5B^{1/2}-1.8\right)$ meV, for $\nu =- 1.$}
\label{figure1}
\end{center}
\end{figure}

\end{widetext}

\section{Summary and discussion}\label{summary}

In summary, motivated by the puzzling results of recent experiments~\cite{Dean11,Feldman12,Feldman13}, we studied fractional and integer quantum Hall states in the zeroth LL of monolayer graphene. We argued that the behavior at filling factors $-2<\nu< 1$ and $-1<\nu<0$ is markedly different, owing to the different importance of short-ranged valley-anisotropic interactions. Qualitatively, one should think about fractional states at $-2<\nu<-1$ as electrons added to vacuum (empty Landau level). Since correlations in fractional states are such that the probability to find two electrons at the same point is negligible, the short-range valley-anisotropic terms do not play any role, which leads to an SU(2)-symmetric effective Hamiltonian. In contrast, fractional states at $-1<\nu<0$ should be viewed as resulting from adding holes to the $\nu=0$ state. Even though holes stay away from each other, they still interact with the background imposed by the $\nu=0$ state, which breaks SU(2) symmetry.

The SU(2) valley symmetry at $-2<\nu<-1$ has two main consequences. First, fractional states with odd numerators must occur via spontaneous breaking of SU(2) symmetry.  As a result, there are gapless neutral Goldstone modes that correspond to fluctuations of the valley polarization. The charge excitations are valley skyrmions, which have lower energy than the quasi-electron/quasi-hole excitations. Furthermore, the SU(2) symmetry of the Hamiltonian makes odd-numerator states very susceptible to weak disorder, which destroys the long-range valley order and introduces a certain density of skyrmions and anti-skyrmions in the ground state. In this case, the energy of adding another skyrmion/anti-skrymion is expected to be strongly reduced compared to the clean system. The lower energy skyrmion excitations and strong sensitivity to disorder make the odd-numerator states at $-2<\nu<-1$ very fragile. 

Second, it is likely that at filling factors with {\it even} numerators in the interval $-2<\nu<-1$ ground states are $SU(2)$-symmetric valley-singlets. In these states the lowest energy charged excitations are quasi-electrons/quasi-holes which can have either valley index, but are spin-aligned, and skyrmions, if they exist, would be energetically costly due to Zeeman energy. The valley-singlet states are not sensitive to weak valley symmetry-breaking disorder (since they cannot take advantage of adjusting valley polarization according to the local disorder configuration, unlike odd-numerator states). Thus, even-numerator states are expected to be quite robust. 

In the interval of filling factors $-1<\nu<0$, the valley and spin symmetries are broken by valley anisotropies and Zeeman interaction, which makes both odd- and even-numerator states robust, and leads to multiple phase transitions between states with different degree of spin and valley polarization. The $\nu=0$ state is a canted AF state, in which states $|Ks\ra$, $|K's'\ra$ with nearly opposite spins $s,s'$ in the $xy$-plane, slightly canted in the $z$-direction, are occupied. The $\nu=-1/3$ state is a partially valley- and spin-polarized Laughlin state of holes $|K\downarrow\ra$ (or $|K'\downarrow\ra$) formed on top of a state with $s=\uparrow,s'=\downarrow$, which optimizes the combination of valley anisotropy and Zeeman energy.

At $\nu=-2/5$ and $-2/3$ there are two competing states: a canted AF state with equal valley occupation (which can be thought of as a singlet state of holes on top of $\nu=0$ state), and a partially polarized state with unequal valley and spin occupations. While anisotropy energy is very close in these competing states, the polarized state can take advantage of the spin Zeeman energy by aligning spin direction of the majority component with the magnetic field. At low magnetic fields, Zeeman energy is negligible, and the states with equal occupation of two valleys, favored by the Coulomb interactions, are expected to be ground states. At higher fields, when Zeeman interaction becomes increasingly important compared to the difference of Coulomb energies of the competing states, a phase transition into a partially polarized ground state is expected. Similarly, at other filling factors at $-1<\nu<0$, we expect phase transitions between composite fermion states with different degree of spin and valley polarization.

By contrast, no phase transitions are expected at filling factors $\nu=-4/3$ or $8/5$, because these states are valley-singlets and are spin-polarized along magnetic field such that they take maximum advantage of the Zeeman energy. At filling factors $\nu=-14/9,-11/7$, and $-7/5$, phase transitions are in principle possible: at low $B$, minimally spin and valley-polarized states are favored by the Coulomb interactions (these correspond to filling lowest possible CF levels), while at high $B$ maximally spin-polarized states will be favored by the Zeeman interaction. 

The above picture is consistent with the recent experiments~\cite{Dean11,Feldman12,Feldman13}. First, odd-numerator states at $-2<\nu< -1$ were either completely absent or very weak in both compressibility~\cite{Feldman12,Feldman13} and transport measurements~\cite{Dean11}. This is likely due to effects of disorder which greatly increases compressibility. The robust even-numerator singlets have been seen~\cite{Feldman12}, and no phase transitions as a function of field have been observed at this point. Second, robust odd and even-numerator states have been seen at $-1<\nu<0$~\cite{Feldman12,Feldman13}, and a series of phase transitions as a function of magnetic field have been observed at all filling factors except for $\nu=-1/3$. 

Turning to integer states, we argued that at  $\nu = 0$, the elementary charged excitations could be a single electron or hole, or could be a small skyrmion, which would have a similar energy cost.    At $\nu=  -1$, we argued that the negative charged   excitation could be a large area valley skyrmion or  a small area spin skyrmion,  while the positive (hole-like) excitation should be a large area valley skyrmion, which would reduce the size of the gap relative to that at $\nu=0$.   Effects of disorder should  further reduce the measured energy gap at $\nu= -1$, relative to that at $\nu=0$.  This might explain the observed differences in the chemical potential jumps at these two filling factors, which are shown in data reported in this paper.

Although we have assumed above that the parameters $g_\perp$ and $g_z$ fall in the range where $\nu=0$ state is in the AF phase, our principal conclusions will be unchanged if the ground state were the Kekule phase or the CDW phase. The main difference is that there would be no canted states in these cases. The states for $0<|\nu|<1$ would always have a majority spin parallel to $B$ and a minority spin in the opposite direction, if the populations are unequal. 

Throughout the paper we have assumed the valley sensitive interactions to
be instantaneous in time.  For phonon-mediated interactions, this is
correct to the extent that the phonon frequency is large compared to the
Coulomb exchange energy within a Landau level.  Retardation effects could
lead to corrections to our result that the energy of states with $|\nu |
\leq 1$ should be independent of the orientation of the valley
polarization $<\vec{\tau}>$.

\section{Future directions}\label{outlook}

Further insights into the nature of fractional states in the zeroth LL of graphene can be obtained from experiments with parallel magnetic field. At a fixed perpendicular magnetic field, it should be possible to observe phase transitions at various filling factors in the interval $-1<\nu<0$ as a function of parallel field component. Moreover, the dependence of the energy gaps on the parallel field should reveal the spin structure of excitations at different filling factors.  

In the future, it would be also interesting to study the fractional quantum Hall effect in higher LLs of monolayer graphene. Owing to the different structure of LL wave functions, the effective interactions will be different compared to the zeroth LL in graphene and to GaAs. This will change the phase diagram of fractional states, as well as their energy gaps. Valley anisotropies in higher LLs in monolayer graphene will be different from the zeroth LL, which may completely change the valley and spin structure of the fractional states. 

Furthermore, new effective interaction regimes are realized in related materials, such as bilayer graphene, which will almost certainly lead to interesting phenomena in the fractional quantum Hall effect regime. Theoretical studies suggest that band structure tunability in bilayer graphene may allow one to tune effective electron interactions by applying perpendicular electric field~\cite{Papic12}. This can be used to stabilize the desired fractional states and induce phase transitions between them. Additional tunability can be achieved by changing the dielectric environment of the sample~\cite{Papic11}. 

Finally, another promising direction is to study the structure of the FQH edge states in graphene. Owing to the atomically sharp confinement~\cite{Li12}, it is possible to avoid edge reconstruction~\cite{Hu11}. A recent theoretical study~\cite{Hu11} suggests that this might enable the observation of universal Luttinger liquid behavior at the edges of FQH states.

 
\section*{Acknowledgments}

The authors have benefited  from an exchange of ideas with Maxim Kharitonov, who  has also studied this problem, and has independently arrived at a number of similar conclusions. 
We also thank Rudolf Morf for enlightening discussions and sharing his unpublished results, and Andrea Young and Pablo Jarillo-Herrero for sharing their data on the nature of $\nu=0$ state in graphene prior to publication. DA thanks Zlatko Papic for helpful discussions. 

This work is supported by the US Department of Energy, Office of Basic Energy Sciences,
Division of Materials Sciences and Engineering under Award No. DE-SC0001819. This work was performed in part at the Center for Nanoscale Systems (CNS), a member of the National Nanotechnology Infrastructure Network, which is supported by the National Science Foundation under NSF award no. ECS-0335765. CNS is part of Harvard University

\end{document}